\documentclass[conference]{IEEEtran}
\IEEEoverridecommandlockouts
\usepackage{cite}
\usepackage{amsmath,amssymb,amsfonts}
\usepackage{algorithmic}
\usepackage{graphicx}
\usepackage{textcomp}
\usepackage{xcolor}
\def\BibTeX{{\rm B\kern-.05em{\sc i\kern-.025em b}\kern-.08em
    T\kern-.1667em\lower.7ex\hbox{E}\kern-.125emX}}
\begin{document}
\title{Type III solar radio burst detection and classification: A deep learning approach\\
}

\author{\IEEEauthorblockN{ Jeremiah Scully}
\IEEEauthorblockA{\textit{ Dept. of Computer and Software Engineering} \\
\textit{Athlone Institute of Technology}\\
Athlone, Ireland \\
j.scully@research.ait.ie}
\and
\IEEEauthorblockN{ Ronan Flynn}
\IEEEauthorblockA{\textit{ Dept. of Computer and Software Engineering} \\
\textit{Athlone Institute of Technology}\\
Athlone, Ireland \\
rflynn@ait.ie}
\and
\IEEEauthorblockN{ Eoin Carley}
\IEEEauthorblockA{\textit{School of Cosmic Physics} \\
\textit{Dublin Institute of Advanced Studies}\\
Dublin, Ireland \\
eoin.carley@dias.ie}
\and
\IEEEauthorblockN{ Peter Gallagher}
\IEEEauthorblockA{\textit{School of Cosmic Physics} \\
\textit{Dublin Institute of Advanced Studies}\\
Dublin, Ireland \\
peter.gallagher@dias.ie}
\and
\IEEEauthorblockN{ Mark Daly}
\IEEEauthorblockA{\textit{ Dept. of Computer and Software Engineering} \\
\textit{Athlone Institute of Technology}\\
Athlone, Ireland \\
mdaly@ait.ie}
}

\maketitle

\begin{abstract}
Solar Radio Bursts (SRBs) are generally observed in dynamic
spectra and have five major spectral classes, labelled Type I to Type V depending on their shape and extent in frequency and time. Due to their complex characterization, a challenge in solar radio physics is the automatic detection and classification of such radio bursts. Classification of SRBs has become fundamental in recent years due to large data rates generated by advanced radio telescopes such as the LOw-Frequency ARray, (LOFAR). Current state-of-the-art algorithms implement the Hough or Radon transform as a means of detecting predefined parametric shapes in images. These algorithms achieve up to 84\% accuracy, depending on the Type of radio burst being classified. Other techniques include procedures that rely on Constant-False-Alarm-Rate detection, which is essentially detection of radio bursts using a de-noising and adaptive threshold in dynamic spectra. It works well for a variety of different Types of radio bursts and achieves an accuracy of up to 70\%. In this research, we are introducing a methodology named You Only Look Once v2 (YOLOv2) for solar radio burst classification. By using Type III simulation methods we can train the algorithm to classify real Type III solar radio bursts in real-time at an accuracy of 82.63\% with a maximum 77 frames per second (fps).
\end{abstract}

\section{Introduction}
Ever since astronomer Richard Christopher Carrington observed the first ever solar flare back in 1859, researchers have been both astonished and perplexed by the behavior of the suns weather and in particular solar flares.

Solar flares are the most energetic explosive phenomena in the solar system, often involving the acceleration of particles to near light-speed\cite{Lin}. The accelerated particles emit light from across the entire electromagnetic spectrum, from gamma rays to radio waves. Radio emission is often high-intensity and is generally observed as complex patterns in dynamic spectra known as solar radio bursts\cite{Pick}. SRBs generally come in five classifications, labelled Type I to Type V, depending on their shape in dynamic spectra. They can occur at a rate of thousands per day (especially Type III bursts), so it is a computational challenge to automatically detect their occurrence and determine their spectral characteristics. However, due to their complex characterisation, classifying these radio bursts is a fundamental challenge. This challenge has been made more complex in recent years with new technology such as LOFAR providing high-volume data streams (up to 3 Gb/s at a single station) of radio burst observations that need to be classified with high speed and accuracy. The necessity of automated data pipelines for solar radio bursts has been made even more apparent in recent years, with the the design of LOFAR for Space Weather (LOFAR4SW)\cite{Lofar4sw}, a system upgrade which aims at autonomously monitoring of solar radio activity (as well as heliospheric and ionospheric activity). Software pipelines to automatically detect SRBs will be an indispensable part of such a system in the near future. This paper establishes the role of machine learning playing a crucial part of such a pipeline.

There have been various attempts to automatically detect SRBs in dynamic spectra in recent years. The current state-of-the-art involve algorithms that implement the Hough or Radon transform as a means of detecting predefined parametric shapes in images\cite{Lobzin}. These algorithms achieve up to 84\% accuracy, depending on the Type of radio burst being classified.
Other techniques include procedures that rely on Constant-False-Alarm-Rate (CRAF) detection\cite{Salmane}, which is essentially detection of radio bursts using a de-noising and adaptive threshold in dynamic spectra. It works well for a variety of different Types of radio bursts and achieves an accuracy of up to 70\%. 

In recent years neural networks have been applied to the problem, in which multi-modal deep learning was applied to a spectrogram at millimetric wavelengths\cite{Ma}. The system used auto-encoders and standard regularization in tandem to achieve a burst detection accuracy of 82\%, however this has not been applied to metric wavelengths (the range of LOFAR) where the bursts can have much more complex shapes. 

Recent research has turned to object detection algorithms such as Faster R-CNN to identify solar radio bursts\cite{Hou}. This deep learning neural network proved to be accurate at extracting small features of solar radio bursts with an average precision (AP) of 91\% however, it doesn't have the performance of real time detection.
One other area in which machine learning is working in tandem with radio telescope observing is SETI (Search for Extra-Terrestrial Intelligence)\cite{Zhang}. SETI use their Allen telescope array to observe planetary systems searching for Fast Radio Bursts (FRBs) using machine learning. SETI highlights regular noise frequency’s and Radio Frequency Interference (RFI) and then isolates irregular high frequency FRB spikes within the dynamic spectrum using a deep Convolutional Neural Network (CNN) called a ResNet\cite{He_Zhang}. This model produced a recall score of 95\%.

With recent research turning to deep CNNs and object detection for classifying and detecting radio frequencies we decided to explore these areas even further. There are many different forms of CNNs for object detection such as You Only Look Once (YOLO)\cite{Redmon_Divvala_Girshick_Farhadi_2016}, Single Shot Detectors\cite{Liu}, Region-CNN (R-CNN)\cite{Girshick}, Fast R-CNN \cite{Girshick_2015}, Faster R-CNN\cite{Ren} and Mask R-CNN\cite{He}. Although the methods mentioned have been proven to be very successful for object detection YOLO has been the only algorithm to offer high accuracy and real time detections on some datasets. 

In this research, we test the accuracy of the deep learning algorithm YOLOv2\cite{Redmon_Farhadi_2017}. using the darkflow framework when applied to Type III solar radio bursts. Using methods to simulate Type III bursts we can create a training set of 80,000 images and train YOLOv2 to identify Type III radio bursts within a dynamic spectrum at an accuracy of 82.63\% at real-time frame rates of above 60fps.\\
In section II of this paper, we will be discussing what LOFAR is and what SRBs are along with the dataset and model configuration used for this research. In section III, we will discuss the training and test set, and also YOLOv2s performance and accuracy on Type III data.

\section{Methodology}
\subsection{LOFAR and I-LOFAR}
LOFAR is a radio interferometer constructed in the north of the Netherlands and across Europe which includes Ireland's very own I-LOFAR station. LOFAR covers the largely unexplored low frequency range from 10–240 MHz and offers a unique number of observing capabilities. LOFAR makes use of Digital beam-forming techniques meaning LOFAR can rapidly re-point it's view to other points of interest in a short space of time. It can also operate simultaneously using multiple stations. LOFAR station can work both in tandem with each other as one whole large radio telescope or singularly as a standalone station. LOFAR antenna stations provide the exact same simple functions as radio dishes of a conventional interferometric radio telescope. Like traditional radio dishes, these stations provide collecting area and raw sensitivity as well as pointing and tracking capabilities. Fundamentally, LOFAR differs from high-frequency radio telescopes in the fact that LOFAR stations do not physically move. LOFAR operates by pointing and tracking using combined signals from the individual antennas to form a phased array using a combination of analog and digital beam forming techniques and in turn makes LOFAR more flexible and agile. Station-level beam-forming allows for rapid re-pointing of the telescope as well as the potential for multiple, simultaneous observations from a given station. The resulting digitized, beam-formed data from the stations can then be streamed to the central processing facility and correlated to produce visibility's for imaging applications and observation analysis. In this research we use the processed observations generated by I-LOFAR as seen in figure 1, to create a test set in which YOLOv2 can be evaluated on.
\begin{figure}[h!]
\includegraphics[width=8.8cm, height=5cm]{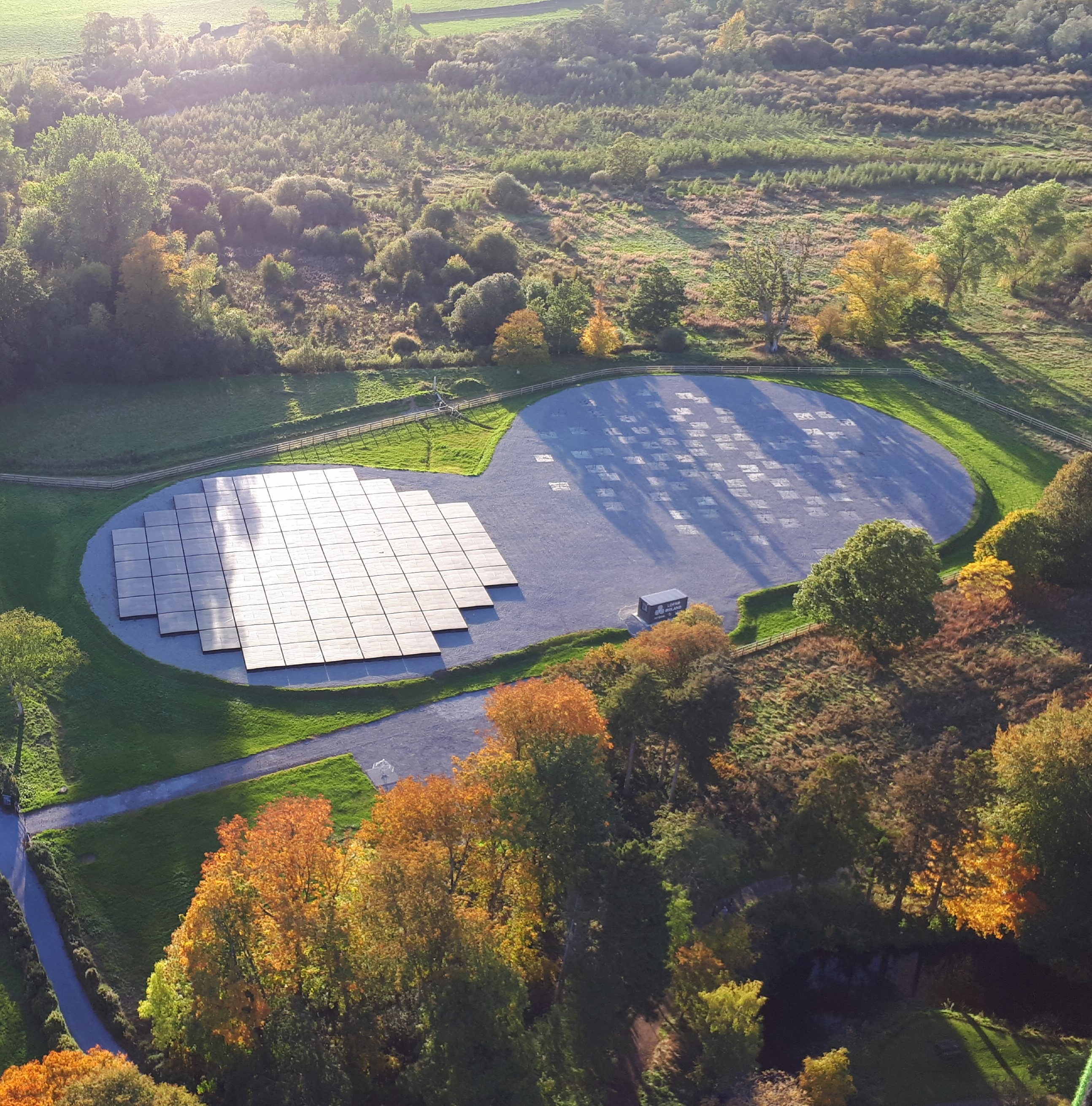}
\caption{I-LOFAR station located in Birr, Co.Offaly, Ireland. This an international station consisting of 96 Low band antennas (right) operating between 10-90MHz and 96 High band antennas (left) operating between 110-250MHz.}
\end{figure}

\subsection{Type III solar radio bursts}
The Sun is an active star that produces large-scale events such as Coronal Mass Ejections (CMEs) and solar flares. Radio emission is often associated with these events in the form of radio bursts. These bursts are classified into five main Types. Type I bursts are short duration narrowband bursts associated with active regions. Type II bursts are slow frequency drifting radio emissions thought to be excited by shock waves travelling through the solar corona and they are associated with CMEs. Type III radio bursts are rapid frequency drifting bursts which can sometimes be followed by continuum emissions; these emissions are called Type V radio bursts. Type IV bursts are broad continuum emissions with rapidly varying time structures. Solar radio bursts are most often observed in dynamic spectra of frequency versus time.

\begin{figure}[h!]
\includegraphics[width=8.8cm, height=5cm]{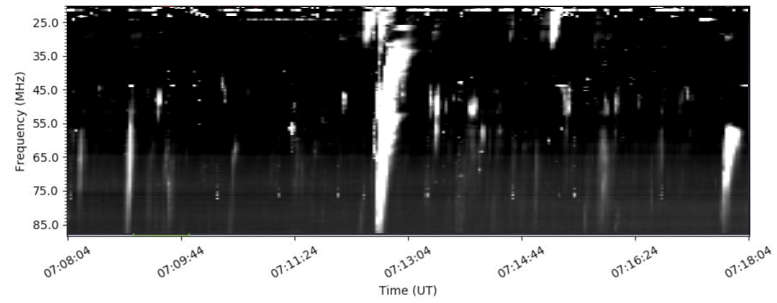}
\caption{Examples of real Type III solar radio bursts as seen in a dynamic spectrum at a frequency range of 20-90MHz. This is used to in the test set to evaluate YOLOv2s performance.}
\end{figure}

The most commonly occurring radio burst is the Type III, which is generally short lasting a couple of seconds and structurally represent a vertical bright strip in dynamic spectra at a frequency range of 10-100MHz, see Figure 2. Although the structure of the Type III seems to be very basic, identifying them is a complex task as they come in a variety of forms within the dynamic spectra. For example, the Type III may be smooth or patchy, weak or strong, superimposed on other radio bursts, standalone or in groups, or may be embedded in strong radio frequency interference (RFI).

\subsection{Dataset}
In order to train a useable model with YOLOv2, a large training dataset is required. For our experiments, this dataset consisted of simulated Type III data created using parametric models. Using these parametric models, we produce Type III radio bursts that are random in number, grouping, intensity, drift rate, and homogeneity. They also allow random variations in the frequency range and duration's of the bursts. We embed the bursts in a background of simulated and random RFI channels, an example which can be found in Figure 3.
\begin{figure}[h!]
\includegraphics[width=8.8cm, height=5cm]{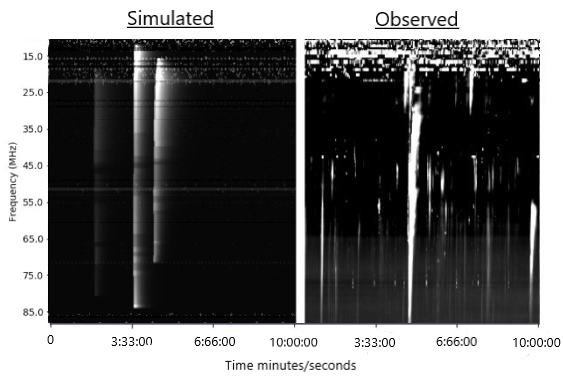}
\caption{A comparison between simulated Type III radio bursts (left) and actual Type III radio bursts (right). The simulated Type III bursts are used to train the deep learning model YOLOv2.}
\end{figure}

While producing these simulated Type III images we needed to label the bounding box coordinates where the simulated Type III was randomly fixed into the image. Using these simulated Type IIIs, a training set of 80,000 simulated Type III SRBs was created to train the YOLOv2 model.
\subsection{Model Configuration}
Once the dataset was created, we need to configure the model. We used a framework called darkflow\cite{Trieu}; a Tensorflow implementation of darknet to develop the model in YOLOv2.
\begin{table}[h!]
\caption{The CNN architecture of YOLOv2. The CNN in YOLOv2 is altered in the fully connected layers of the CNN. The fully connected layers are removed and the detection and classifications are done by K-means classification for improved accuracy.}
\begin{tabular}{ |p{1.9cm}||p{1.6cm}|p{1.8cm}|p{1.8cm}|  }
 \hline
 \multicolumn{4}{|c|}{Darknet-19 Architecture} \\
 \hline
Type& Filters &Size/Stride&Output\\
 \hline
 Convolutional   & 32    &3 X 3&   224 X 224\\
 Maxpool   &     &2 X 2/2&   112 X 112\\
 Convolutional   & 64    &3 X 3&   112 X 112\\
 Maxpool   &     &2 X 2/2&   56 X 56\\
 Convolutional   & 128    &3 X 3&   56 X 56\\
 Convolutional   & 64    &1 X 1&   56 X 56\\
 Convolutional   & 128    &3 X 3&   56 X 56\\
 Maxpool   &     &2 X 2/2&   28 X 28\\
 Convolutional   & 256    &3 X 3&   28 X 28\\
 Convolutional   & 128    &1 X 1&   28 X 28\\
 Convolutional   & 256    &3 X 3&   28 X 28\\
 Maxpool   &     &2 X 2/2&   14 X 14\\
 Convolutional   & 512    &3 X 3&   14 X 14\\
 Convolutional   & 256    &1 X 1&   14 X 14\\
 Convolutional   & 512    &3 X 3&   14 X 14\\
 Convolutional   & 256    &1 X 1&   14 X 14\\
 Convolutional   & 512    &3 X 3&   14 X 14\\
 Maxpool   &     &2 X 2/2&   7 X 7\\
 Convolutional   & 1024    &3 X 3&   7 X 7\\
 Convolutional   & 512    &1 X 1&   7 X 7\\
 Convolutional   & 1024    &3 X 3&   7 X 7\\
 Convolutional   & 512    &1 X 1&   7 X 7\\
 Convolutional   & 1024    &3 X 3&   7 X 7\\
 \hline
 Convolutional   & 1000    &3 X 3&   7 X 7\\
 Avgpool   &     &Global&   1000\\
 Softmax   &     &&   \\
 \hline
\end{tabular}\\
\end{table}
In YOLOv2 there are 19 convolutional layers and 5 maxpool layers as seen in table 1. We needed to change the number of filters in the last convolutional layer as we are only looking to detect one class being Type IIIs. To do this, we use the function (1)
\begin{equation}
filters= bounding * (classes+coords)
\end{equation}
where \(filters\) is the number of filters in the last convolutional layer, \(bounding\) is the number of bounding boxes per grid cell, \(classes\) is the number of classes being detected, and \(coords\) is the number of coordinates in each bounding box.  Here \(bounding=5\), \(classes=1\), and  \(coords=5\). The total number of filters needed \(filters\) at the final convolutional layer is 5 * (1+5)=30.\\
We then altered the sizes of the bounding boxes using the anchor values. We decided to set the bounding box width to be very narrow, rational being Type IIIs are short lasting in terms of times often a couple of seconds. We then set the height of the bounding box to be the height of 10-90MHz similarly seen in the training set and dynamic spectrum. It's apparent, Type III intensity's drift so low it cannot be seen by the naked eye as seen in Figure 4.
\begin{figure}[h!]
\includegraphics[width=8.8cm, height=4cm]{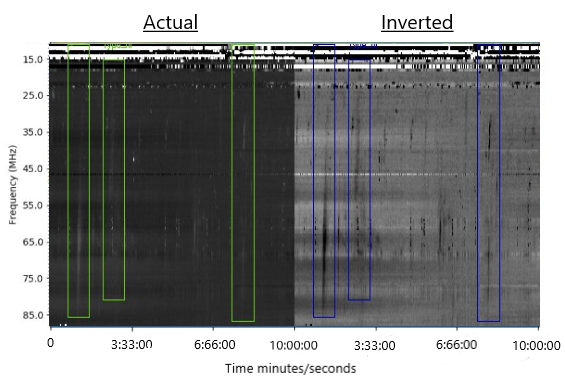}
\caption{An Example case of a Type III intensity fading to the point of being unidentifiable by the naked eye. Once the colours have been inverted the drift of intensity can be seen.}
\end{figure}

In order to achieve real time frame rates with YOLOv2, we needed to decrease the number of bounding boxes predicted on each test image from 845 to 405. To do this, we needed to change the input size from 416x416 to 288x288. YOLOv2’s convolutional layers downsample the image by a factor of 32 so by using an input image of 416/32 we get an output feature map of 13 × 13 or sectioning an image into a 13x13 square grid. This multiplied by the number of bounding boxes per grid cell gives us the number of predictions per image 5x13x13=845. In our case, we are feeding in a 288x288 input image. As a result we get a feature map of 9x9 with 5 bounding boxes per square grid giving us 5x9x9=405 predictions per image. At 288×288 it runs at a maximum 77 fps with accuracy comparable to Fast R-CNN. This input size complements the dataset input into the YOLOv2 model as only one class in greyscale format is being identified.
\section{Results}

\subsection{Training}\label{AA}
The YOLOv2 model is trained to detect and classify Type III SRBs. The training set consists of 80,000 simulated Type III images that are random in number, grouping, intensity, drift rate, in-homogeneity, start-end frequency and start-end time. The training set is also simultaneously labelled as it's created, allowing us to create a high volume training set along with a list of text files containing bounding box coordinates. These text files are then translated into XML to fit the darkflows training set requirements. Once prerequisites have been complete the training set can be fed into the CNN for training. Darkflow, by default, only uses 1 GPU or CPU for training so we had to add the ability for Nvidia Scalable Link Interface (SLI) support to the framework. The model is trained with a learning rate of 0.001, a momentum of 0.9 and Leaky Rectified Linear Unit (ReLU) as an activation function during training. Learning parameters are also updated until convergence using Stochastic Gradient Descent (SGD).

This research has been performed on a machine comprising of 2 x SLI inter-connected GPU Nvidia Geforce RTX 2080 Ti, using Ubuntu 20.4.2 LTS on an AMD Ryzen Threadripper 1950x with 32GB of RAM.  For the training configuration we are use 90\% of GPU capacity for 200 epochs at a batch size of 32. With this configuration, training took 4 days with loss decreasing with every iteration as seen in Figure 5.
\begin{figure}[h!]
\includegraphics[width=8.7cm, height=5cm]{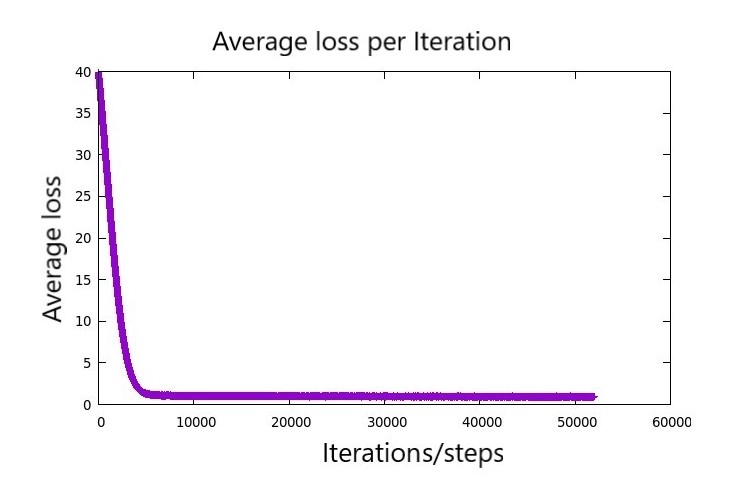}
\caption{A plot showing loss decreasing with every iteration of training. Loss is another method of evaluating how well YOLOv2 models the training dataset.}
\end{figure}

\subsection{Test Set}
The test set for the model is an 12 hour observation made by I-LOFAR on the 10th of September 2017. The raw data is processed and converted into an image and plotted frequency in MHz from 0 to 250 over the duration of the observation (time) as seen in figure 6. This image is then converted to greyscale to match the characteristics of the training set so that colour isn't an influence on the models predictions. We focus on the 10-90 MHz range in the observations as this shows the Type IIIs most prominent attributes. The observation is then cut into 10 minute chunks to create a test set of 1331 images, containing about 15,000 Type III solar radio bursts.
Once we have our test set images, we needed to annotate our ground truth bounding box values. This was done by using LabelImg, an annotating tool used to label objects within an image. Once a Type III is labelled it's corresponding bounding box coordinates is stored in an XML file that will be used to compare the ground truth coordinates to the models predicted coordinates.
\begin{figure}[h!]
\includegraphics[width=8.7cm, height=5cm]{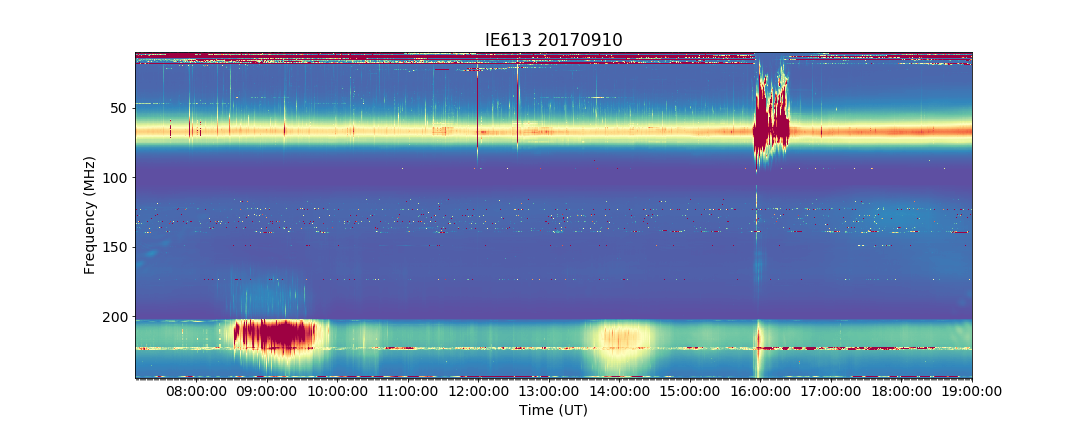}
\caption{An I-LOFAR observation made on the 10th of September 2017. This raw data generated by I-LOFAR has been processed and plotted on a frequency range of 0-250MHz.}
\end{figure}
\subsection{Model Evaluation}
The YOLOv2 model's performance was measured using the test set of 1331 images from I-LOFAR described previously. The unit in which we represent our models performance result is the f1-score. The f1-score highlights the balance between precision and recall with precision meaning how good the model predicted the location of an object and recall measuring how good the model found and located all objects. We use the following functions to calculate the f1-score of the model.
\begin{equation}
Precision = \frac{TP}{TP+FP}
\end{equation}
\begin{equation}
Recall = \frac{TP}{TP+FN}
\end{equation}
\begin{equation}
f1-score =  2 * \frac{Precision * Recall}{Precision + Recall}
\end{equation}
True positive (TP) and false positive (FP) are obtained from intersection over union (IoU) from tested data. IoU compares the predicted bounding box to the ground truth bounding box. A prediction is classified as TP if the IoU is greater than 0.5, FP is if it's less than 0.5. A False Negative (FN) is specified for those images where the model failed to detect a known Type III object. One important factor to take into account when evaluating the model performance is confidence threshold.  Confidence threshold measures how confident the model is at predicting a certain object, in this case Type III SRBs. The lower the confidence, the more detections made on a test image but also the more false detections being made. After experimenting with different thresholds it was found that the model was at it's most optimized with confidence threshold set to 0.1, see table 2. With this configuration, the resulting f1-score is 82.63\% for detecting Type III solar radio bursts.

\begin{table}[h!]
\caption{From table 2 we can see that if we use higher threshold the resulting accuracy will be lower. But if we lower the threshold to allow more detections per test image the accuracy will be higher. However, in doing this we can see an increase in false positives also.}
\begin{tabular}{ |p{1cm}||p{1cm}|p{1cm}|p{1cm}|p{1cm}|p{1cm}|  }
 \hline
 \multicolumn{6}{|c|}{Model accuracy per confidence threshold} \\
 \hline
Conf. thresh&Precision& Recall &f1-score&True Positive&False Positive\\
 \hline
 0.5   & 10.79\%    &98.35\%&   19.44\%&1310&22\\
 0.4   &  26.22\%   &98.36\%&   41.41\%&3185&53\\
 0.3   & 46.33\%    &98.13\%&   63.04\%&5628&107\\
 0.25   & 56.81\%    &97.57\%&   63.04\%&6910&172\\
 0.2   & 68.20\%    &96.75\%&   80.01\%&8313&279\\
 0.15   & 75.13\%    &90.73\%&   82.30\%&9220&942\\
 0.1   & 83.42\%    &81.85\%&   82.63\%&9796&1947\\
 \hline
\end{tabular}\\
\end{table}
\section*{Conclusion}
In this paper, we have shown that YOLOv2 is very good at detecting and classifying Type III SRBs at high frame rates. This particular configuration of YOLOv2 can achieve an accuracy of 82.63\% on a real data observation consisting of almost 15,000 Type III solar radio burst examples while also achieving real-time frame rates (maximum 77 fps). This is a significant step towards having an automated solar instrument that offers real time analysis. While current state of the art algorithms such as Hough and randon transform and CRAF offer excellent accuracy, these algorithms don't have the benefit of providing both accuracy and real time performance. Key to attaining the high accuracy with the YOLO model is the quality of the training dataset which is made up entirely of simulated Type III SRBs. We intend to create a new training dataset with real Type III observations included with the simulations. This will be used to test the robustness of the training set. Issues associated with robustness would be the exclusion of embedded RFI and general background noise as well as compensating for intensity fluctuations in real data.  In conclusion, we have shown that accurate real-time classification of Type III SRBs is readily attainable.

\bibliographystyle{ieeetr}
\bibliography{./bibliography/ref.bib}

\end{document}